\documentclass[longauth]{aa} 

\usepackage{graphicx}
\usepackage{txfonts}
\begin{document}

   \title{A geometric distance measurement to the\\ Galactic Center black hole with 0.3\% uncertainty}


   \author{The GRAVITY  Collaboration: R.~Abuter\inst{8}, A.~Amorim\inst{6}, M.~Baub\"ock\inst{1}, J.P.~Berger\inst{5}, H.~Bonnet\inst{8}, W.~Brandner\inst{3}, Y.~Cl\'enet\inst{2}, V.~Coud\'e~du~Foresto\inst{2}, P.T.~de~Zeeuw\inst{10,1}, J.~Dexter\inst{1}, G.~Duvert\inst{5}, A.~Eckart\inst{4,13}, F.~Eisenhauer\inst{1}, N.M.~F\"orster Schreiber\inst{1}, P.~Garcia\inst{7}, F.~Gao\inst{1}, E.~Gendron\inst{2}, R.~Genzel\inst{1,11}, O.~Gerhard\inst{1}, S.~Gillessen\inst{1}, M.~Habibi\inst{1}, X.~Haubois\inst{9}, T.~Henning\inst{3}, S.~Hippler\inst{3}, M.~Horrobin\inst{4}, A.~Jim\'enez-Rosales\inst{1}, L.~Jocou\inst{5}, P.~Kervella\inst{2}, S.~Lacour\inst{2,1}, V.~Lapeyr\`ere\inst{2}, J.-B.~Le Bouquin\inst{5}, P.~L\'ena\inst{2}, T.~Ott\inst{1}, T.~Paumard\inst{2}, K.~Perraut\inst{5}, G.~Perrin\inst{2}, O.~Pfuhl\inst{1}, S.~Rabien\inst{1}, G.~Rodriguez Coira\inst{2}, G.~Rousset\inst{2}, S.~Scheithauer\inst{3}, A.~Sternberg\inst{12,14}, O.~Straub\inst{1}, C.~Straubmeier\inst{4}, E.~Sturm\inst{1}, L.J.~Tacconi\inst{1}, F.~Vincent\inst{2}, S.~von Fellenberg\inst{1}, I.~Waisberg\inst{1}, F.~Widmann\inst{1}, E.~Wieprecht\inst{1}, E.~Wiezorrek\inst{1}, J.~Woillez\inst{8} \& S.~Yazici\inst{1,4}
   \fnmsep\thanks{GRAVITY has been developed by a collaboration of the Max Planck Institute for Extraterrestrial Physics, LESIA of Paris Observatory / CNRS / UPMC / Univ. Paris Diderot and IPAG of Universit\'e Grenoble Alpes / CNRS, the Max Planck Institute for Astronomy, the University of Cologne, the Centro de Astrof\'isica e Gravita{\c{c}}\^ao, and the European Southern Observatory.}
          }

\institute{Max Planck Institute for Extraterrestrial Physics (MPE), Giessenbachstr.1, 85748 Garching, Germany
\and
LESIA, Observatoire de Paris, PSL Research University, CNRS, Sorbonne Universit\'es, UPMC Univ. Paris 06, Univ. Paris Diderot, Sorbonne Paris Cit\'e, 92195 Meudon Cedex, France
\and
Max-Planck-Institute for Astronomy, K\"onigstuhl 17, 69117 Heidelberg, Germany
\and
1. Physikalisches Institut, Universit\"at zu K\"oln, Z\"ulpicher Str. 77, 50937 K\"oln, Germany
\and
Univ. Grenoble Alpes, CNRS, IPAG, 38000 Grenoble, France
\and
CENTRA and Universidade de Lisboa - Faculdade de Ci\^encias, Campo Grande, 1749-016 Lisboa, Portugal
\and
CENTRA and Universidade do Porto - Faculdade de Engenharia, 4200-465 Porto, Portugal
\and
European Southern Observatory, Karl-Schwarzschild-Str. 2, 85748 Garching, Germany
\and
European Southern Observatory, Casilla 19001, Santiago 19, Chile
\and
Sterrewacht Leiden, Leiden University, Postbus 9513, 2300 RA Leiden, The Netherlands
\and
Departments of Physics and Astronomy, Le Conte Hall, University of California, Berkeley, CA 94720, USA
\and
School of Physics and Astronomy, Tel Aviv University, Tel Aviv 69978, Israel
\and
Max-Planck-Institute for Radio Astronomy, Auf dem H\"ugel 69, 53121 Bonn, Germany
\and
Center for Computational Astrophysics, Flatiron Institute, 162 5th Ave., New York, NY, 10010, USA
}


 
 \abstract{We present a 0.16\% precise and 0.27\% accurate determination of $R_0$, the distance to the Galactic Center. Our measurement uses the star S2 on its 16-year orbit around the massive black hole Sgr~A* that we followed astrometrically and spectroscopically for 27 years. Since 2017, we added near-infrared interferometry with the VLTI beam combiner GRAVITY, yielding a direct measurement of the separation vector between S2 and Sgr~A* with an accuracy as good as $20\,\mu$as in the best cases. S2 passed the pericenter of its highly eccentric orbit in May 2018, and we followed the passage with dense sampling throughout the year. Together with our spectroscopy, in the best cases with an error of $7\,$km/s, this yields a geometric distance estimate: $R_0  = 8178 \pm 13_\mathrm{stat.} \pm 22_\mathrm{sys.}\,$pc. This work updates our previous publication in which we reported the first detection of the gravitational redshift in the S2 data. The redshift term is now detected with a significance level of $20\sigma$ with $f_\mathrm{redshift} = 1.04 \pm 0.05$.}
 
 \authorrunning{GRAVITY collaboration} 

   \maketitle
  %

\section{Introduction}

Measuring distances is a key challenge in astronomy. While many distance estimators rely on secondary calibration methods, there are a few methods that are the basis for the whole distance ladder. These methods all compare an angular scale on sky with a size known in absolute terms. Foremost is of course the parallax method. It compares an observed reflex motion on the sky, measured in angular units with the size of Earth's orbit. Recently, GAIA improved the number and quality of parallaxes available substantially (GAIA collaboration 2018). However, GAIA working in the optical and at moderate spatial resolution does not provide any parallaxes towards the crowded and highly dust obscured center of the Milky Way. The extinction can be overcome by observing at longer wavelengths, in the near-infrared (NIR, $1-5\,\mu$m). Very large telescopes with adaptive optics, and recently interferometry between large telescopes (GRAVITY collaboration 2017), overcome the stellar crowding. This allowed us to determine the orbits of 40 stars around the central massive black hole with periods between 13 and a few thousand years (Gillessen et al. 2017). These stars offer another direct method of determining a distance. The distance to the Galactic Center (GC), $R_0$, can be determined by comparing the radial velocities (measured in km/s) of these stars with their proper motions (measured in mas/yr). The measurement is direct, since this can be done for individual stellar orbits, as opposed to using a sample of stars together with a dynamical model like in \citet{2006A&A...445..513V} for the globular cluster $\omega$~Cen or in \citet{2015MNRAS.447..952C}  for the Milky way nuclear cluster.
  
Most suitable for the orbit method is the star S2 on a 16-year orbit (the second shortest period known so far, \citet{2012Sci...338...84M}), with a semi-major axis $a \approx 125\,$mas. S2 has an apparent K-band magnitude of $m_K \approx 14$, bright enough for spectroscopy. It is a massive, young main sequence B star \citep{2003ApJ...586L.127G, 2008ApJ...672L.119M, 2017ApJ...847..120H} offering a few atmospheric absorption lines in the observable parts of the spectrum. Several works used S2 to measure the distance to the GC. The first measurement was in \citet{2003ApJ...597L.121E} who reported $R_0 = 7940 \pm 420\,$pc. \citet{2005ApJ...628..246E} updated this value to $R_0 = 7620 \pm 320\,$pc. \citet{2008ApJ...689.1044G} reported $R_0 = 8400 \pm 400\,$pc, \citet{2009ApJ...692.1075G} $R_0 = 8330 \pm 350\,$pc. More recently, \citet{2016ApJ...830...17B} measured $R_0 = 7860 \pm 140 \pm 40\,$pc, and \citet{2017ApJ...837...30G} obtained $R_0 = 8320 \pm 70 \pm 140\,$pc. Here and in what follows, the first error is statistical, and the second is systematic. All these measurements rely on adaptive optics data. For general, recent overviews of $R_0$ determinations see \citet{2010RvMP...82.3121G} and \citet{2016ARA&A..54..529B}.
  
S2 passed the pericenter of its orbit in May 2018, an event that we followed in detail both with astrometry and spectroscopy \citep{2018A&A...615L..15G}. The primary goal of these observational efforts was the detection of relativistic effects in the orbital motion. However, the data also allow for an unprecedentedly accurate measurement of $R_0$, because of the large swing in radial velocity (from +4000 to -2000 km/s) and the large orbital phase covered in 2018. In \citet{2018A&A...615L..15G} we presented the detection of the gravitational redshift from Sgr~A* in the S2 spectra. Our previous analysis included data up to end of June 2018. It addressed the question, whether the gravitational redshift and Doppler terms are in agreement with the predictions of Einstein's theory of relativity. At the same time, our orbital solution also included the most precise determination of $R_0$ so far: $R_0= 8122 \pm 31\,$pc, where the error is statistical only. Several authors studying the Milky Way structure used this result already \citep{2018RNAAS...2c.156M, 2018RNAAS...2d.210D, 2019ApJ...870L..10M, 2019ApJ...871..120E}. Here, we update our value for $R_0$, using data up to the end of 2018, and we simply apply the relativistic corrections assuming General Relativity is correct, yielding one fit parameter less. Also, we investigate the systematic error on $R_0$ from our measurement, which we did not consider in \citet{2018A&A...615L..15G}.
 
\section{Data}

In \citet{2018A&A...615L..15G}, we used 45 AO-based astrometric points (after down-sampling), 77 radial velocities, and 30 GRAVITY interferometric data points. The present study adds ten epochs of radial velocity measurements from late June 2018 to late September 2018, and ten epochs of GRAVITY astrometry. Furthe, we re-analyzed our radial velocity data from SINFONI and the GRAVITY astrometry, implementing an improved understanding of the respective systematic effects. This also led to a slightly different data selection, and different grouping of the observations.

For the SINFONI data we re-visited the wavelength calibration, yielding an improved wavelength dispersion solution. Where possible we determined the radial velocities by template fitting. The uncertainties are a combination of 
formal fit error, wavelength error and the error introduced by selecting a certain extraction mask in the field of the integral field unit. For the details see appendix~\ref{appa}.

For the GRAVITY data we replaced the manual frame selection with an objective outlier-rejection, and included the (minor) effect of atmospheric differential refraction. The data analysis includes data selection, binary fitting, correction for atmospheric refraction, outlier rejection, nightly averaging, correction for effective wavelength, adding systematic errors, and error scaling. We report the details in appendix~\ref{appb}. 

Overall, our new data set consists of 169 adaptive optics (AO) based astrometry points between 1992 and 2019, 91 radial velocities between 2000 and 2019, and 41 GRAVITY-based astrometry points in 2017 and 2018. 

Our adaptive optics data set samples the on-sky motion of S2 at high cadence. The distance between subsequent data points is typically smaller than the size of the point spread function. Any confusion event with unrecognized faint stars thus might affect several data points, leading to correlated measurements. As in \citet{2018A&A...615L..15G} we therefore down-sampled the AO data set into intervals of constant arc length on the sky, and we down-weighted these AO data by a factor two, in order to take into account the additional uncertainty due to unseen confusion events. Further, we omitted the 2018 data where additional confusion with Sgr~A* affects the data, leading to 48 AO-based astrometric data points. Further, we developed a different approach for the same issue, namely a noise model (see sec.~\ref{sec3}). This gives a second data set, in which we use all 169 AO-based astrometric points.

   \begin{figure*}
   \centering
  \includegraphics[width=0.75\linewidth]{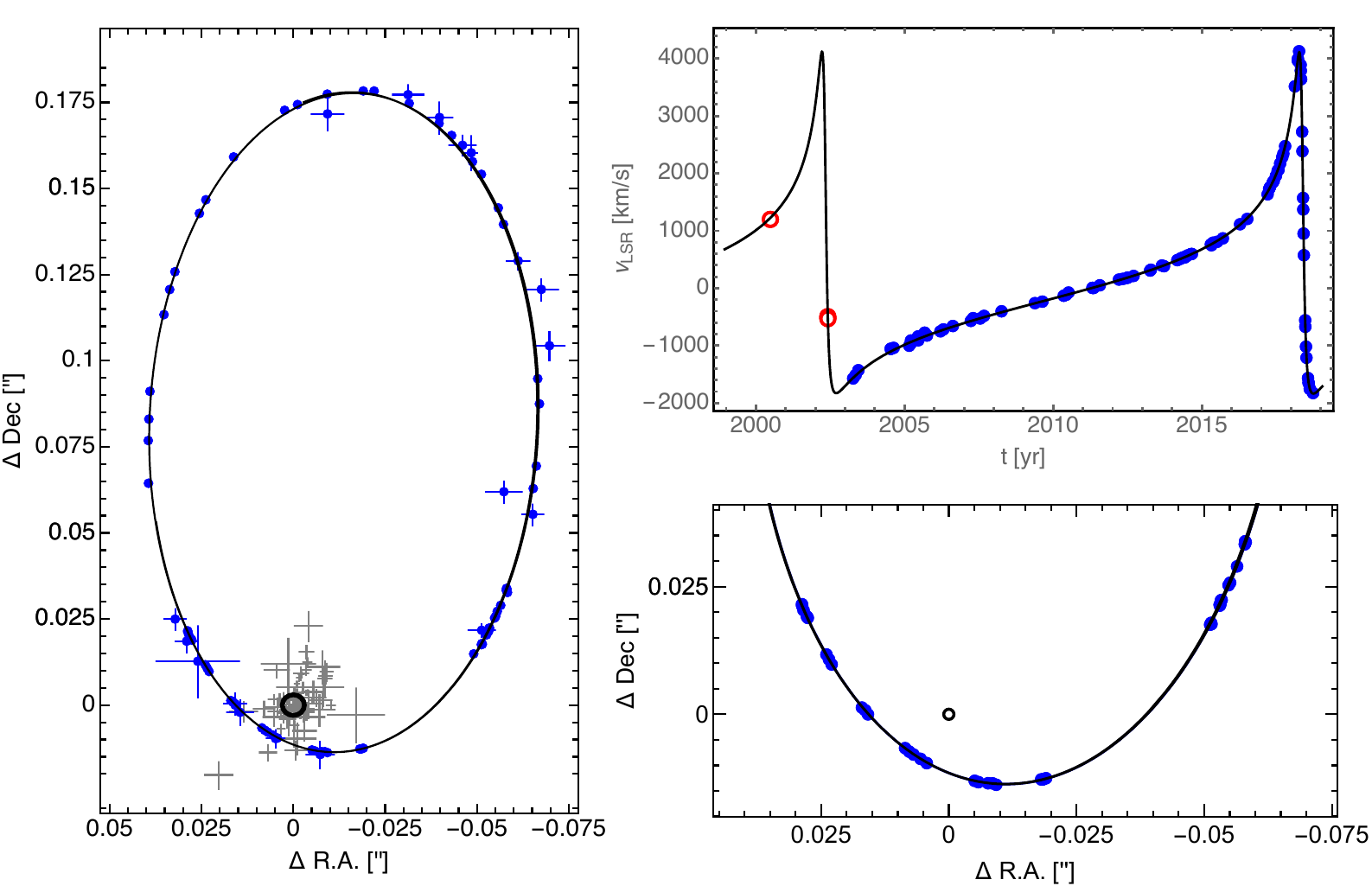}
     \caption{The orbit of S2. Left: On-sky view of the astrometric data (blue) in the down-sampled version with the best-fit orbit (black ellipse). The black circle marks the position of Sgr~A*. The locations of previous, AO-based flares agree with that position (gray crosses). Right top: The radial velocity data of S2 together with the best-fit orbit. The blue data are from the VLT, the red are earlier epochs from the Keck data set \citep{2008ApJ...689.1044G}. Right bottom: Zoom into the on-sky orbit in 2017 and 2018, showing the GRAVITY data that have error bars smaller than the symbol size.}
     \label{fig1}
    \end{figure*}

\section{Analysis}
\label{sec3}
We used the same techniques as in \citet{2018A&A...615L..15G} and \citet{2017ApJ...837...30G}. The analysis essentially consists of one step: determining the best-fit orbit for the data given, and the corresponding uncertainties. We employed a $\chi^2$-minimization for finding the best-fit, and for the uncertainties we used the standard error matrix approach, a Markov chain Monte Carlo (MCMC) sampler and a bootstrapping technique. The latter bootstraps an artificial data set by drawing from the original data, separately for the AO astrometry, the radial velocities and the GRAVITY data. In order to avoid issues from the AO points being correlated, we used the down-sampled data set for the bootstrapping.

For a different approach with the AO data, we implemented a noise model of the type presented in \citet{2018MNRAS.476.4372P} for the AO-based astrometry. Such a model has the advantage that it estimates the additional amount of error and the correlation length from the data themselves, avoiding any prior choices on how to treat the data. In our implementation, we exchanged the temporal correlation length of \citet{2018MNRAS.476.4372P} with a spatial one. The underlying source of correlation is confusion with unseen stars, which one can describe naturally by a length scale in the image plane. Since S2 has a widely varying proper motion, a temporal correlation length is less suited. This model adds two additional parameters to fit for: The spatial correlation length and the typical confusion amplitude - which correspond to the down-sampling and down-weighting in the other data set. We note that fitting the noise model is feasible only when we also use the GRAVITY data - otherwise its parameters are too degenerate with the other 13 parameters. We did not exclude all 2018 data for this data set, but only the epochs at which Sgr~A* apparently affected the position measurements, as visible in an elongated source structure or excess flux of S2.

We also analyzed a third data set excluding all AO astrometry. Perhaps somewhat surprisingly, the two years of GRAVITY data already are the much stronger constraint for the orbit compared to the past 27 years of AO imaging data. 

Compared to the analysis in \citet{2018A&A...615L..15G}, we included in the calculation of the transverse Doppler effect the apparent proper motion of Sgr~A* to the South West of $(-3.151, -5.547)\,$mas/yr, a reflex of the solar motion around the Milky Way center \citep{2004ApJ...616..872R}. This corresponds to $v_\odot \approx 250\,$km/s, while S2 at pericenter reaches an on-sky motion of $v_\mathrm{S2} \approx 7320\,$km/s. Since in the Doppler formula a term of type $(v_\mathrm{S2} + v_\odot)^2 \approx v_\mathrm{S2}^2 (1 + 2 v_\odot/v_\mathrm{S2})$ occurs, the proper motion of Sgr~A* leads to a small, but noticeable correction. The change in the parameter $f_\mathrm{redshift}$, by which we measured the strength of the redshift term in \citet{2018A&A...615L..15G} is $\Delta f_\mathrm{redshift} = + 0.038$, and the change in distance is $\Delta R_0 = +6\,$pc.

\section{Results}
\subsection{The distance $R_0$ to the Galactic Center}
Having as few free parameters in the fit as possible yields the most precise estimate for $R_0$. Hence we assumed General Relativity holds and fix the parameter $f_\mathrm{redshift}$, by which we measured the strength of the redshift term in \citet{2018A&A...615L..15G}, to $f_\mathrm{redshift} = 1$. We further used the R{\o}mer delay and included the first order correction from the Schwarzschild metric. The coordinate system parameters only apply to the AO astrometry, since GRAVITY directly measures the vector S2~-~Sgr~A*.

   \begin{figure*}
   \centering
  \includegraphics[width=0.75\linewidth]{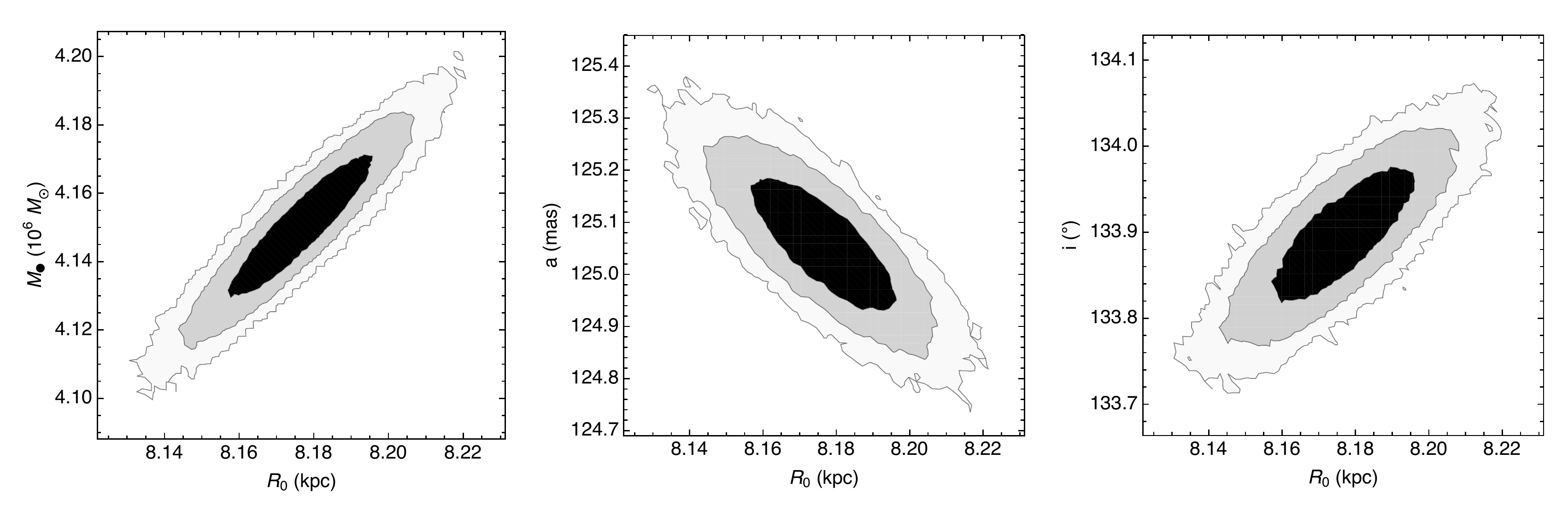}
     \caption{Selected posterior densities as obtained from the MCMC sampler with N=200000, here for the noise model data set. The contour lines mark the 1-, 2-, and 3.$\sigma$ levels. We only show the diagrams with the strongest correlations. All parameters are well determined \citep{2018A&A...615L..15G}.}
     \label{fig2}
    \end{figure*}

We list our best-fit results in Table~\ref{tab1}, and show the best fit in Fig.~\ref{fig1}. The error bars reported are the formal fit errors from the error matrix. The three data sets yield completely consistent parameters within the formal uncertainties. The reduced $\chi^2$ values by construction of the errors are close to 1 (appendices~\ref{appa}~\&~\ref{appb}). 

\begin{table*}
\caption{Best-fit parameters for our three data sets. The parameters $x_0$, $y_0$, $vx_0$, $vy_0$ describe the location and motion of the mass in the coordinate system of the AO data in R.A. and Dec. Since GRAVITY directly measures the separation vector, we do not need to include such coordinate system offsets for the GRAVITY data. The third velocity, $vz_0$ is the offset of the motion in the radial direction along the line of sight, the negative sign means a blue-shift or a motion towards the observer. The parameters $(a, e, i, \Omega, \omega, t_\mathrm{peri})$ are the classical orbital elements semi-major axis, eccentricity, inclination, position angle of ascending node, longitude of pericenter, and the epoch of pericenter passage. The orbital elements are defined as the osculating orbital elements at $t = 2010.0$, i.e. the conversion to position and velocity is done at that epoch assuming a Kepler orbit.}             
\label{tab1}      
\centering          
\tiny{
\begin{tabular}{l | c c c  }     
{\bf Parameter} & {\bf down-sampled data} & {\bf noise model fit} & {\bf GRAVITY only} \\      
\hline
$R_0$ [pc] & $8179 \pm 13$ & $8178 \pm 13$ & $8175 \pm 13$ \\    
mass [$10^6\,\mathrm{M}_\odot$] & $4.154 \pm 0.014$ & $4.152 \pm 0.014$ & $4.148 \pm 0.014$ \\    
$x_0$ [mas] & $-1.04 \pm 0.36$ & $-0.65 \pm 0.36$ & N.A. \\    
$y_0$ [mas] & $-0.47 \pm 0.35$ & $-0.73 \pm 0.35$ & N.A. \\    
$vx_0$ [$\mu$as/yr] & $68 \pm 31$ & $68 \pm 32$ & N.A. \\    
$vy_0$ [$\mu$as/yr] & $158 \pm 31$ & $108 \pm 31$ & N.A. \\    
$vz_0$ [km/s] & $-3.3 \pm 1.5$ & $-3.0 \pm 1.5$ & $-2.8 \pm 1.5$ \\    
$a$ [mas] & $125.072 \pm 0.084$ & $125.066 \pm 0.084$ & $125.065 \pm 0.086$ \\    
$e$ & $0.884282 \pm 0.000064$ &  $0.884293 \pm 0.000064$ & $0.884288 \pm 0.000064$ \\    
$i$ [$^\circ$] & $133.911 \pm 0.052$ &$133.904 \pm 0.052$ & $133.883 \pm 0.053$ \\    
$\Omega$ [$^\circ$] & $228.067 \pm 0.041$ & $228.075 \pm 0.041$  & $228.091 \pm 0.041$  \\    
$\omega$ [$^\circ$] & $66.250 \pm 0.035$ & $66.253 \pm 0.035$ & $66.257 \pm 0.035$\\    
$t_\mathrm{peri}$ [yr] - 2018& $0.3790 \pm 0.0014$ & $0.3790 \pm 0.0014$ & $0.3789 \pm 0.0014$ \\    
UTC date & 19.5.2018 09:53 & 19.5.2018 09:51  & 19.5.2018 09:47  \\    
red. $\chi^2$ & 0.82 &1.10 & 1.00 \\    
\end{tabular}
}
\end{table*}

   \begin{figure}
   \centering
  \includegraphics[width=0.47\linewidth]{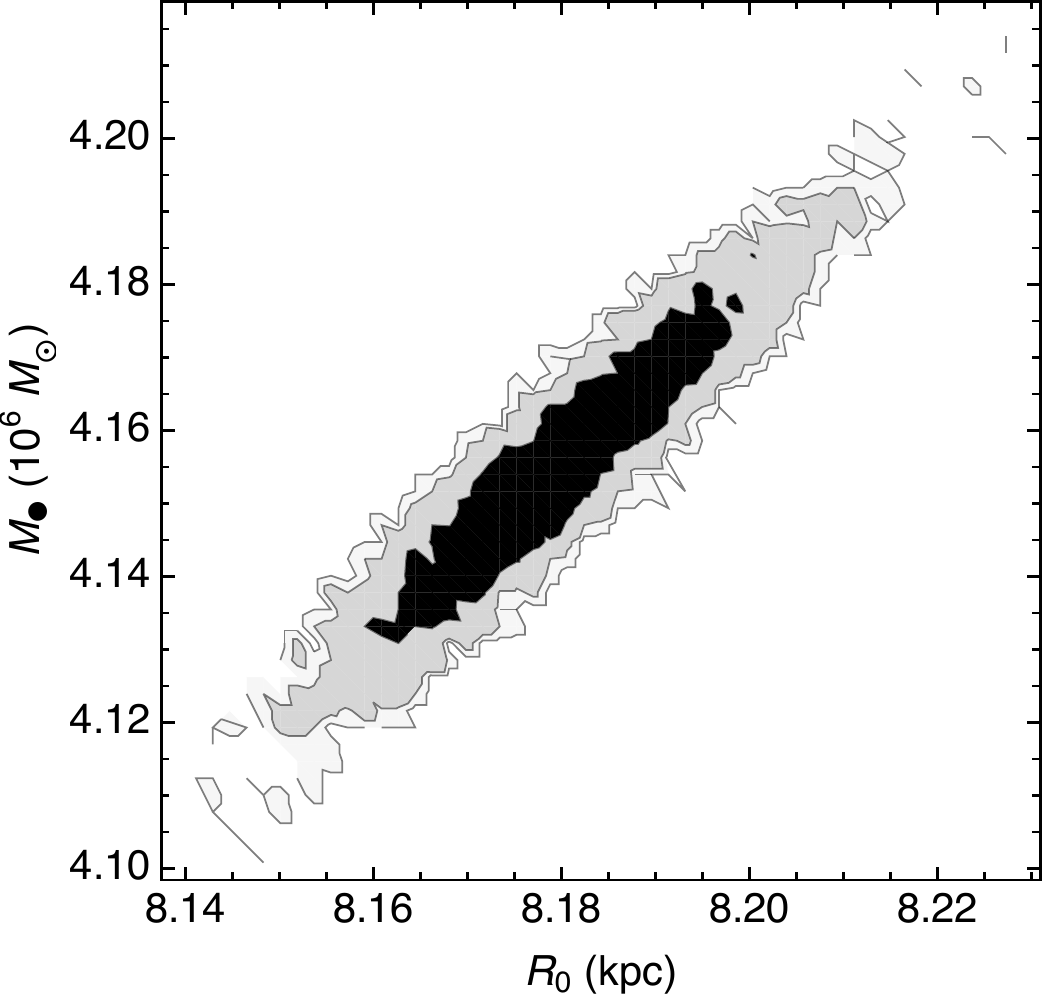}
     \caption{Posterior distribution for $R_0$ and mass from our bootstrap sample. The contour lines mark the 1-, 2-, and 3.$\sigma$ levels.}
     \label{fig3}
    \end{figure}

The noise model has two additional free parameters, the noise amplitude $\sigma = 0.83 \pm 0.15\,$mas and the spatial correlation length $\lambda = 21.2 \pm 3.8\,$mas. These numbers define by how much a certain data point is expected to be off from the model, given the other data. The correlation length is of the same order of magnitude as the AO point spread function radius, and the amplitude is reasonable. Our best-fit $\sigma$ corresponds to a perturbing star of $m_K \approx 17$ at a distance of our best-fit $\lambda$ \citep{2018MNRAS.476.4372P}.

   \begin{figure}
   \centering
  \includegraphics[width=0.8\linewidth]{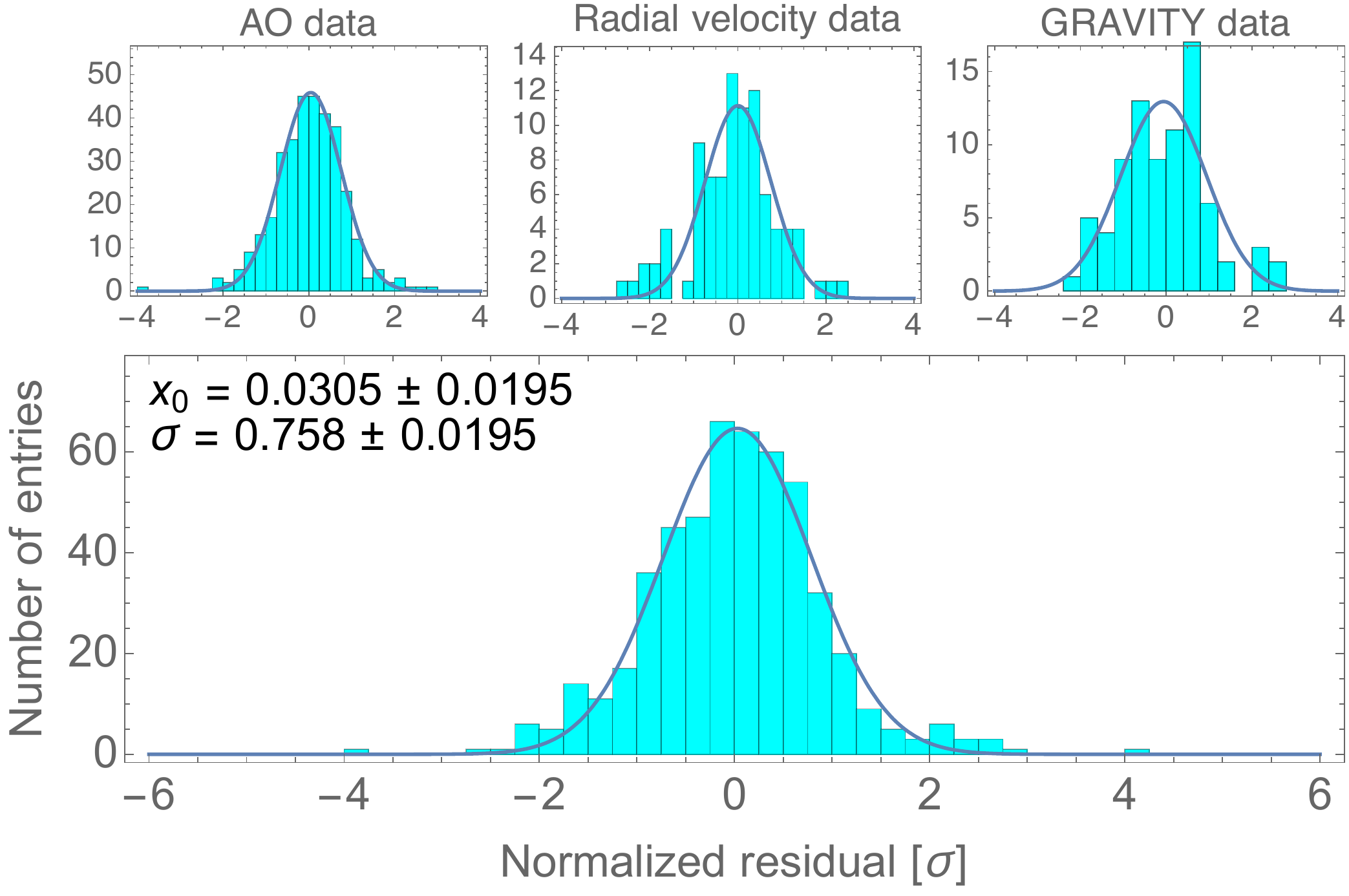}
     \caption{Histograms of the normalized residuals, the ratio of residual to error for each data point. Top row: Individually for the three subsets of data. Bottom: The combined data set.}
     \label{fig3a}
    \end{figure}

Using the MCMC sampler, we obtained the full 13-dimensional posterior distribution. All parameters are well constrained, and Fig.~\ref{fig2} shows the diagrams with the strongest parameter correlations: Mass-vs-$R_0$, semi-major axis-vs-$R_0$, and inclination-vs-$R_0$. The most probable value agrees with the best-fit value, and the 1-$\sigma$ uncertainty from the posterior is $13\,$pc, fully consistent with the estimate from the error matrix.
We further estimated our errors by bootstrapping (and re-fitting each artificial data set). For this, we used the down-sampled data set, since here the most important correlation between data points is removed. Fig.~\ref{fig3} shows the resulting distribution for N = 5000 bootstraps. The most likely value agrees with the best-fit value, and so do the error bars: $R_0 = 8178^{+13}_{-12}\,$pc.

Fig.~\ref{fig3a} shows the normalized residual (residual divided by the error) distributions for each of the three subsets of data and for the whole data set. The distributions are well-behaved and reasonably close to a Gaussian with mean 0 and width 1.
      
The size of the $R_0$ error of $13 / 8178 \approx 0.16\%$ is comparable to what a simple estimate yields. $R_0$ is directly related to the ratio of proper motion (arc length divided by time) and radial velocity. The most constraining part of the orbit is the pericenter swing, which we followed with GRAVITY in 2017 and 2018. 
\begin{itemize}
\item The arc length is $\approx 150\,$mas, more than $1000\times$ larger than the median, 2D error of the 41 GRAVITY points\footnote{The median 1D error of the 2018 GRAVITY data is $60\,\mu$as, for 2017 it is $145\,\mu$as. These numbers already take into account the scatter from night to night. The uncertainties for individual data points within a single night are smaller \citep{2018A&A...618L..10G}. The difference in median error between 2017 and 2018 is caused by the improvement in fiber positioning implemented for 2018. The median error over the whole data set of 41 points is $86\,\mu$as 1D or equivalent $121\,\mu$as 2D.}. The astrometric precision is thus at the 0.01\% level and does not contribute significantly to the statistical error. 
\item The median error of the radial velocity data in 2017 and 2018 is $14.4\,$km/s, and we have 35 data points. The mean absolute radial velocity of our data in 2017 and 2018 is $2300\,$km/s. The spectroscopic precision 
is thus at the 0.1\% level. It dominates the measurement error, and it is of the same magnitude as the actual statistical error on $R_0$.
\end{itemize}
We conclude that $R_0 = 8178 \pm 13_\mathrm{stat.}\,$pc. However, we still lack an estimate for the systematic error.

\subsection{Systematic errors}
Our estimate for $R_0$ is direct and as such does not depend on intermediate calibration steps. Any systematic error is directly related to how accurately we understand the instruments we use, i.e. how accurate are the on-sky positions we measure and how accurate are the radial velocities. 
Fig.~\ref{fig2} shows the strongest parameter correlations for $R_0$ from the posterior distribution of the 13-dimensional fit. They are with mass, semi-major axis and inclination. These correlations can be understood qualitatively.
\begin{itemize}
\item $R_0$ is inversely proportional to the semi-major axis $a$. A biased determination of $a$ in angular units would bias $R_0$, since the radial velocity data determine $a$ in absolute units - for S2 $a \approx 1023\,$AU. The slope of the correlation in Fig.~\ref{fig2} (middle) confirms this, $R_0 \times a \approx 1023\,$AU. The instrumental reason why $a$ could be biased is an error in the image scale. A scale error of 1\% would imply a distance error of $\approx 80\,$pc.
\item The inclination $i$ would be biased, if the image scale were off in one dimension only. The MCMC shows a sensitivity of $R_0$ to $i$ of $3.75^\circ/$kpc. At the inclination of S2, the sensitivity of the scale change to a change in $i$ amounts to 1.2\%/$\,^\circ$. 
\item Kepler's third law, $G M = 4\pi^2 (a \times R_0)^3/P^2$ (where the semi-major axis $a$ is measured in angular units), shows that our mass measurement is equivalent to determining the period $P$, since the nominator $a \times R_0$
is a constant, see above. The MCMC shows a sensitivity of $R_0$ to $M$ of $1.4 \times 10^3\,M_\odot/$pc at the best fit $R_0$, corresponding to $\approx 1 \mathrm{day}\,$/ pc for the sensitivity to $P$. Note that the error we make in measuring the period due to the uncertainty in the underlying data is captured in the statistical error on $R_0$. What matters here would be a systematic error in measuring time, which we can exclude at the relevant level. The mass-distance degeneracy is not a source of potential systematic error. 
\end{itemize}

We conclude that if the parameter degeneracies were to introduce a systematic error on $R_0$, it would originate from an error in the astrometry. Further, we note that the GRAVITY data completely dominate our astrometry (see table~\ref{tab1}), and that the AO + GRAVITY data sets yield the same result as the GRAVITY-only fit. Hence the uncertainty in the GRAVITY astrometry dominates the systematic error from the astrometry. In appendix~\ref{appc} we show that we estimate this uncertainty to $19\,$pc or 0.24\%.

When using the GRAVITY astrometry, we assume that the NIR counterpart of Sgr~A* is at the position of the center of mass. In \citet{2018A&A...618L..10G} we discovered that the flaring emission from Sgr~A* moves in a circular pattern with a radius of a few Schwarzschild radii, $\approx50\,\mu\mathrm{as}$. The center of the motion matches the position of the mass also to within $\approx50\,\mu\mathrm{as}$. We use that as uncertainty on our assumption and  estimate the effect on  $R_0$  by artificially displacing the mass by that amount. Doing so to the North, South, East and West yields changes in $R_0$ of $+8\,$pc, $-8\,$pc, $-6\,$pc, $+5\,$pc. We include $6\,$pc in the systematic error for the assumption that GRAVITY directly measures the separation vector between S2 and mass center.

  \begin{figure}
   \centering
  \includegraphics[width=0.47\linewidth]{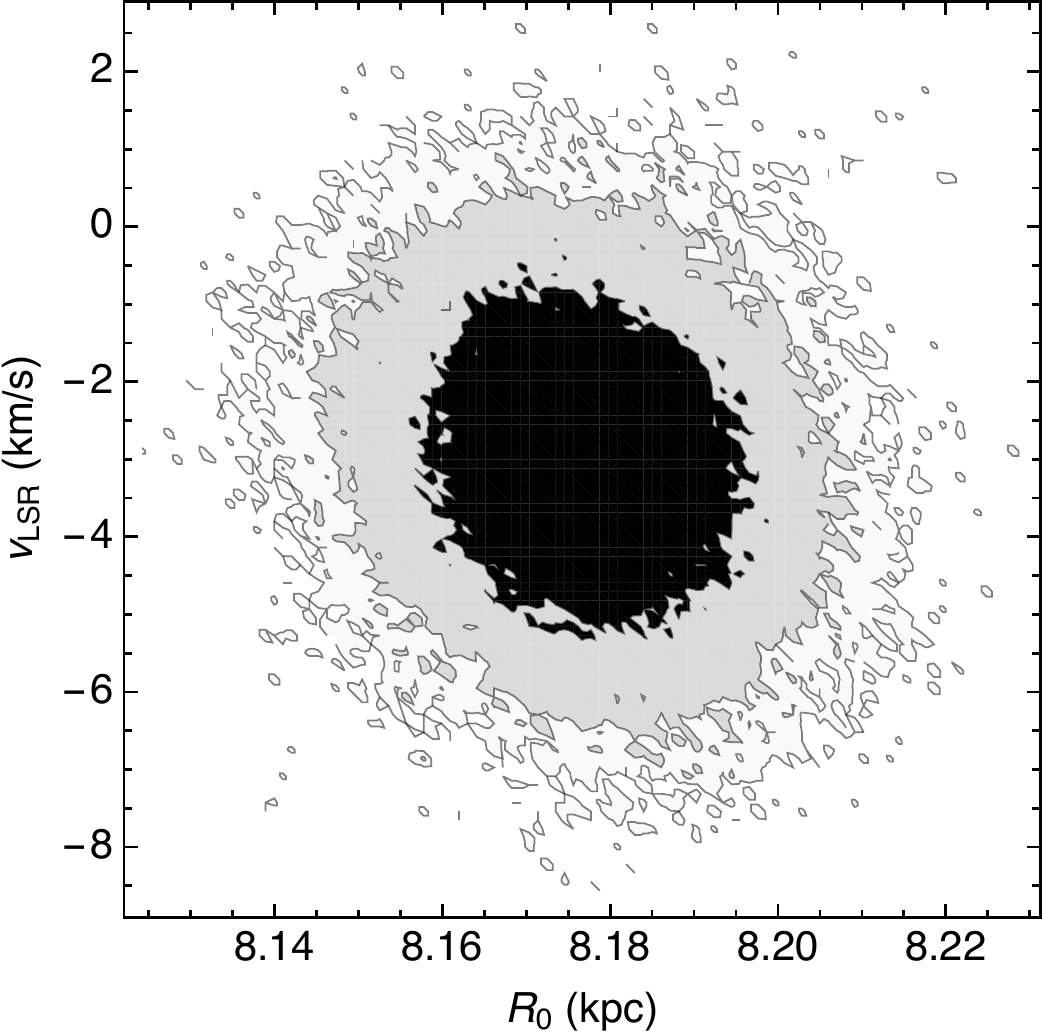}
     \caption{Posterior distribution for $R_0$ and the offset in radial velocity. The contour lines mark the 1-, 2-, and 3.$\sigma$ levels.}
     \label{fig4}
    \end{figure}
  
With full coverage of the orbit the measurement of $R_0$ is no longer degenerate with the offset $vz_0$ in radial velocity (Fig.~\ref{fig4}, cf. \citet{2008ApJ...689.1044G}). A general offset in the radial velocity would be absorbed fully into $vz_0$, but it would not affect our measurement of $R_0$. The 0-th order of the wavelength calibration is thus not a source of systematic error, and if one exists, the leading order could only be the first order, i.e. the dispersion solution. 

Our spectra are calibrated with a higher order polynomial, using multiple atmospheric lines in the same spectra as calibration points. From the residuals of our dispersion solution at these calibration points, we estimate the systematic uncertainty in the wavelength axis to $2.5\,$km/s over the range relevant for S2. Together with the mean absolute radial velocity in 2017 and 2018 ($2300\,$km/s), we obtain a systematic error of 0.11\% or $9\,$pc.

Taken together, we estimate thus our systematic error on $R_0$ to $22\,$pc. Our main result is 
\begin{equation}
R_0  = 8178 \pm 13_\mathrm{stat.} \pm 22_\mathrm{sys.}\,\mathrm{pc}.\nonumber
\end{equation}
The statistical error is dominated by the measurement uncertainties of the radial velocities, the systematic error by the GRAVITY astrometry.

\subsection{Update on the gravitational redshift in S2}
\begin{figure*}
   \centering
  \includegraphics[width=0.75\linewidth]{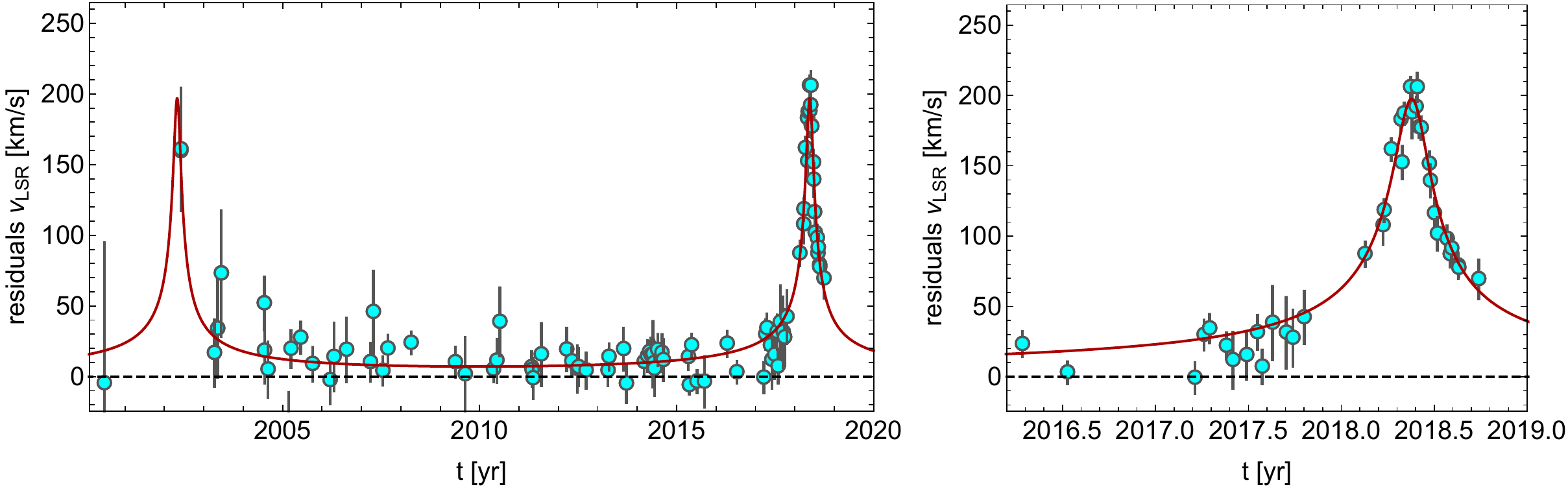}
     \caption{Update of the posterior analysis of \citet{2018A&A...615L..15G}. The panels show the residuals of the radial velocity data to the best-fit orbit in which post-fit the redshift and transverse Doppler effect were turned off (line at 0, $f_\mathrm{redshift} = 0$). The 2018 data show a highly significant excursion. The red line gives the orbit with $f_\mathrm{redshift} = 1$. General relativity is an excellent description for the residuals.}
     \label{fig5}
    \end{figure*}

With the new data sets in hand, we repeated the posterior analysis of \citet{2018A&A...615L..15G} to check for the combined effect of gravitational redshift and transverse Doppler effect. We parameterize the strength of the effects with an artificial parameter $f_\mathrm{redshift}$ such that $f_\mathrm{redshift}=0$ corresponds to classical physics, while $f_\mathrm{redshift}=1$ corresponds to the effects occurring as predicted by General Relativity. Using an orbit model including the first order correction due to the Schwarzschild metric and including the R{\o}mer delay we find  $f_\mathrm{redshift} = 1.047 \pm 0.052$ for the noise model fit and $f_\mathrm{redshift} = 1.036 \pm 0.052$ when using the down-sampled data set. Fig.~\ref{fig5} shows the radial velocity residuals to the classical part of the true best-fit orbit. For this we set $f_\mathrm{redshift} = 0$ without refitting, after having fit with $f_\mathrm{redshift}=1$. We compare these residuals to the true model (i.e. with the effects turned on,$f_\mathrm{redshift}=1$). We exclude that purely Newtonian physics can describe our data at a significance level of $20\sigma$.

\subsection{Distance estimate without radial velocities}
\begin{figure}
   \centering
  \includegraphics[width=0.43\linewidth]{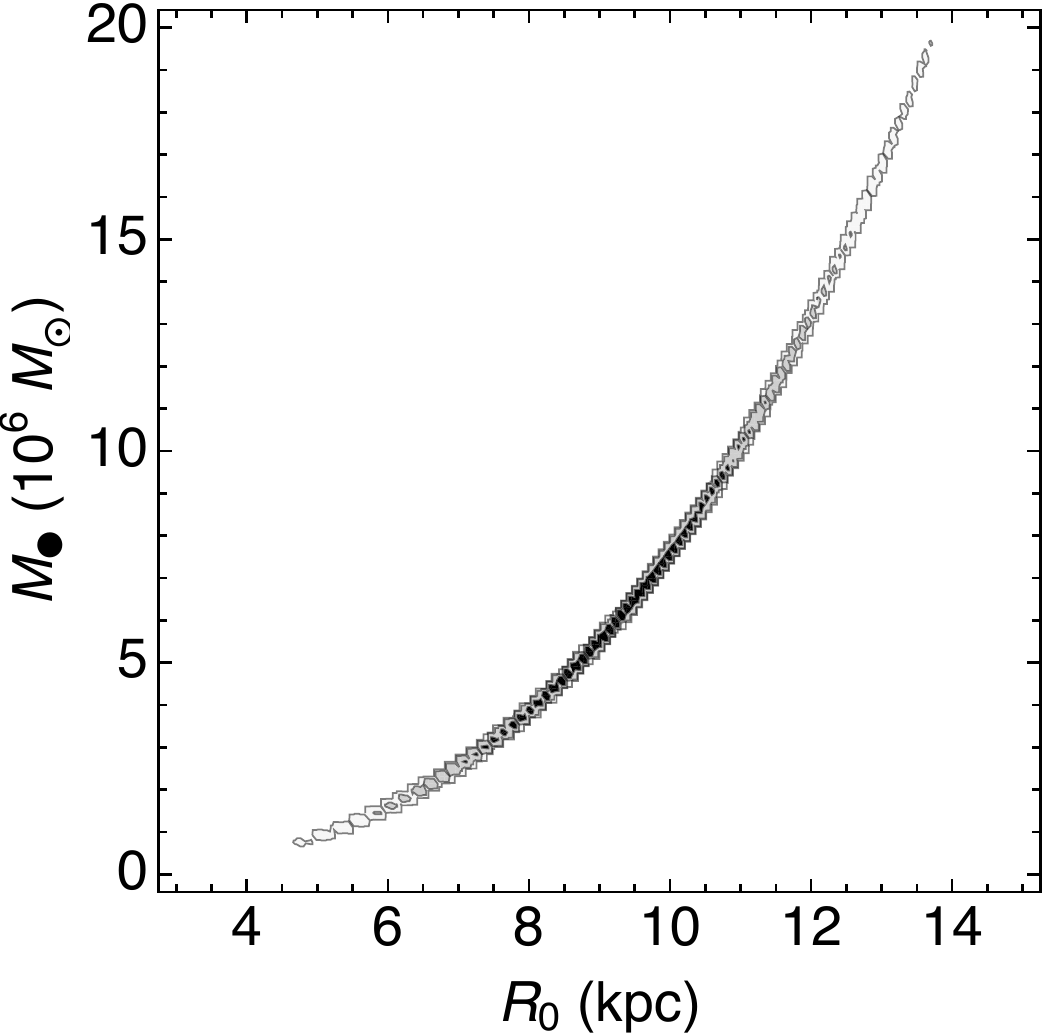}
  \includegraphics[width=0.43\linewidth]{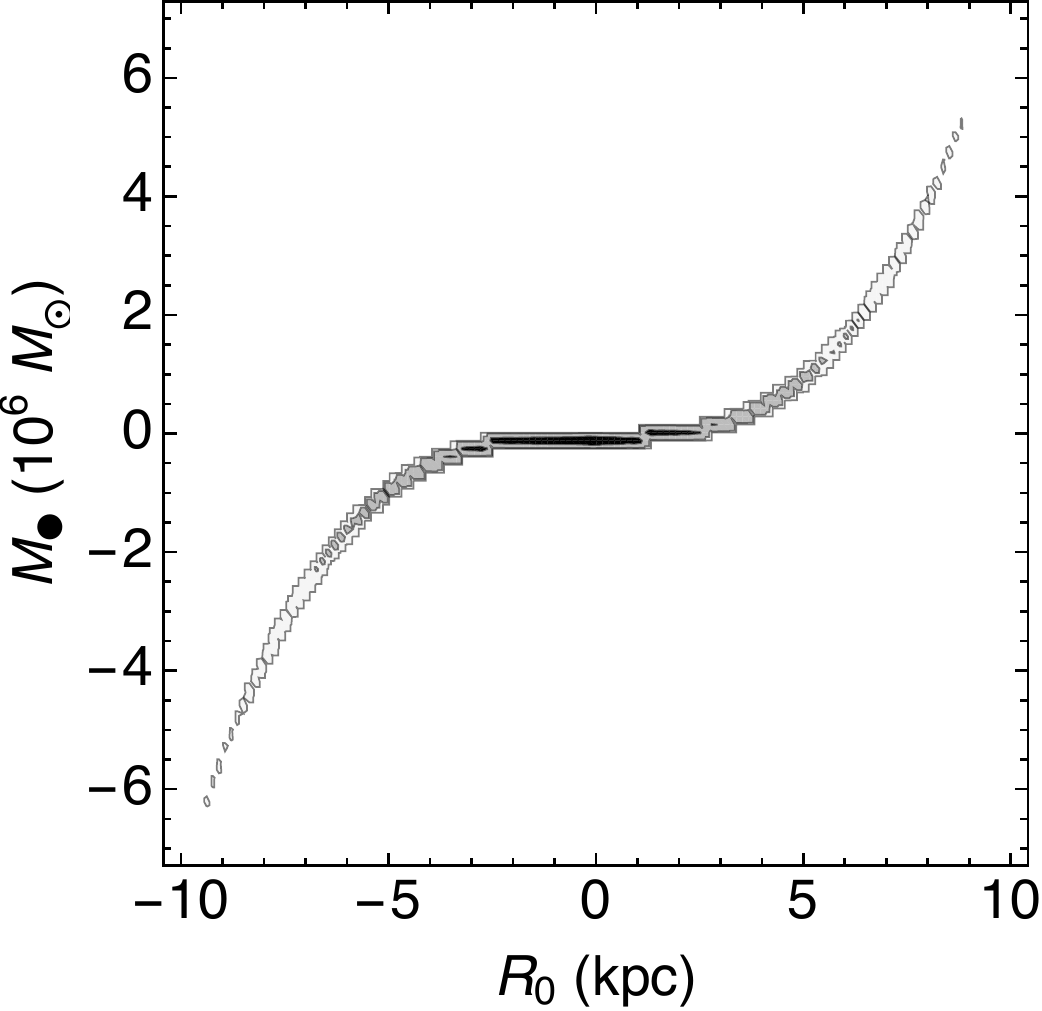}
     \caption{Posterior distributions for the data set without radial velocities. Left: Using the correct orbit model. Right: Using an orbit model that neglects the R{\o}mer effect. In this plot we allowed negative distances (and correspondingly negative masses) to avoid having a bound of the parameter space at 0, where the actual maximum of the distribution falls.}
     \label{fig6}
    \end{figure}

Our GRAVITY measurement provides also the first direct distance measurement from orbital motion without the need for radial velocities. The key for that is the R{\o}mer effect: The light travel time across the orbit makes astrometric points appear a bit ahead or lagging behind the orbit, depending on whether S2 is in front of or behind Sgr~A*. For a Keplerian orbit with astrometric data only, and no light-time travel effect, the distance cannot be determined. The best-fit mass and distance are degenerate along a line $M\propto R_0^3$. Given that the light travel time across the orbit between 2017 and 2018 (where we have GRAVITY data) is of order 3 days, and that we can detect S2's daily motion in the GRAVITY data, our astrometry breaks the degeneracy. Fig.~\ref{fig6} (left) shows that this is indeed the case. The best-fit distance for this case is $R_0 = 9.5 \pm 1.5\,$kpc, consistent with our best estimate. To our knowledge \citet{2006A&A...449.1281A} were the first to propose this type of distance measurement, but we are not aware of an application anywhere so far.

If we were ignoring the R{\o}mer effect for the purely astrometric data set, one does not get back a fully degenerate mass-distance relation. Instead, the fit then tries to get as small a distance as possible (figure 6, right), i.e. in the sense of a limit, one gets $R_0 \rightarrow 0$. This is where the light travel time effect is minimal, as imposed by the wrong orbit model without R{\o}mer delay. This just shows in a different way that our astrometry requires a finite speed of light, and thus can estimate $R_0$.

\section{Discussion}
The best estimate for $R_0$ from \citet{2016ARA&A..54..529B} using only their set of ten independent best measurements that don't invoke Sgr~A* is $R_0 = 8210 \pm 80\,$pc, in perfect agreement with our value. This means that Sgr~A* indeed is at the center of the Milky Way bulge.

Our value of $R_0$ together with the proper motion of Sgr~A* of $6.379 \pm 0.026\,$mas/yr$\,=30.24 \pm 0.12\,$km/s/kpc from \citet{2004ApJ...616..872R} implies $\Theta_0 + V_\odot = 247.4 \pm 1.4\,$km/s, where $\Theta_0$ is the rotation speed of the local standard of rest (LSR) and $V_\odot$ is the peculiar solar motion toward $l = 90^\circ$. The error on $\Theta_0 + V_\odot$ is composed roughly equally of the error in the proper motion of Sgr~A* and the uncertainty in $R_0$. This constraint on $\Theta_0 + V_\odot$ is compatible with the recent determination from \citet{2018ApJ...867L..20H}, who found $\Theta_0 + V_\odot = 253 \pm 6\,$km/s from GAIA astrometry of the Sgr stream.

\citet{2016ARA&A..54..529B} estimate $V_\odot = 11\pm2\,$km/s, but to take into account the radial variations in the median $v_\phi$ seen by \citet{2018A&A...616A..11G} we use a total uncertainty of $4\,$km/s. Together with our estimate for $\Theta_0 + V_\odot$ this implies $\Theta_0 = 236.9 \pm 4.2\,$km/s. From combining GAIA DR2 and APOGEE data \citet{2019ApJ...871..120E} found $\Theta_0 = 229 \pm 6\,$km/s, where the error is the reported systematic uncertainty. \citet{2019MNRAS.485.3296W} used GAIA DR2 and RR Lyrae stars to derive $\Theta_0 = 217 \pm 6\,$km/s. Using trigonometric parallaxes of high-mass star forming regions \citet{2014ApJ...783..130R} find $\Theta_0 = 240 \pm 8\,$km/s. 

Another remarkable result is the fact that the offset in the radial velocity, $vz_0$, is small and consistent with zero. The offset absorbs any possible systematic offset in the radial velocity.
\begin{itemize} 
\item The surface gravity of S2 contributes $\Delta vz_0 = G M_\mathrm{S2} / r_\mathrm{S2} c = 1.6\,$km/s \citep{2003A&A...401.1185L}, where we used $r_\mathrm{S2}$, the radius of S2, and $M_\mathrm{S2}$, its mass, from 
\citet{2017ApJ...847..120H}. 
\item The contribution of the Galactic potential can be approximated by $\Delta vz_0 = v_\odot^2/c \ln (R_0 / R_\mathrm{S2})$, where $v_\odot$ is the Sun's circular galactocentric speed and $R_\mathrm{S2}$ is the galactocentric radius of S2 \citep{2003A&A...401.1185L}. The approximation surely does not hold inside the sphere of influence of Sgr~A* ($\approx 3\,$pc), where the MBH dominates the potential. However, due to the logarithm in the expression, the actual effective value for $R_\mathrm{S2}$ does not matter strongly. With $v_\odot \approx 230\,$km/s and $R_\mathrm{S2} = 3\,$pc one gets $\Delta vz_0 =1.4\,$km/s, using the apocenter distance $R_\mathrm{S2} = 0.009\,$pc, the number is $\Delta vz_0 =2.4\,$km/s.
\item Frame-dragging by a maximally-spinning black hole could have an average $\lesssim0.2\,$km/s contribution to the redshift \citep{2010ApJ...720.1303A, 2017A&A...608A..60G}. 
\item Light bending and Shapiro delay reach $\lesssim 4\,$km/s \citep{2010ApJ...720.1303A} but are highly peaked around pericenter and flip sign, so that they do not induce 
a bias on $vz_0$.
\item Contributions from the solar system are around $3\,$m/s, and thus negligible.
\end{itemize}

A similarly sized offset in $vz_0$ might arise from the uncertainty of the construction of the LSR, that should have by its original definition no motion component in the radial direction, $U_{\rm LSR} = 0$.
The LSR correction applied to our data uses the values from \citet{2010MNRAS.403.1829S}, who report $U_\odot=11.10_{-0.75}^{+0.69}\,$km/s, where $U_\odot$ is the solar motion in the direction of the GC. In their review \citet{2016ARA&A..54..529B} conclude $U_\odot =  10.0 \pm 1.0\,$km/s. The variations in the median radial
velocity of stars measured by \citet{2018A&A...616A..11G} in the nearby disk
suggest that $U_{\rm LSR}$ is uncertain on the scale of several km/s. 

Further, an offset in $vz_0$ could be due to the intrinsic motion of Sgr~A* with respect to the Milky Way. \citet{2004ApJ...616..872R} measured Sgr~A*'s  motion perpendicular to the Galactic plane to $0.4 \pm 0.9\,$km/s. For the third dimension, the motion along the Galactic plane, \citet{2009ApJ...700..137R} report $-7.2 \pm 8.5\,$km/s, and the update in \citet{2014ApJ...783..130R} implies tighter constraints around $2-3\,$km/s. The expected "Brownian motion" of Sgr~A* due to scattering with stars in its vicinity is yet a bit smaller than these limits with $0.2\,$km/s \citep{2002ApJ...572..371C, 2007AJ....133..553M}.

The parameter $vz_0$ is the sum of these offsets, and we measured it to be small. The most likely reason why the sum is small is that the summands are small. Under this hypothesis we conclude that to within few km/s Sgr~A* is at rest at the center of the Milky and that the LSR is moving tangentially. 

Our data constrain very strongly the angular diameter of Sgr~A*. Due to the correlation between mass and $R_0$, the constraint is stronger than what simple error propagation would yield. We find $R_S/R_0 = 10.022 \pm 0.020_\mathrm{stat.}\pm 0.032|_\mathrm{sys.}\,\mu$as. The combined uncertainty corresponds to $50000\,$km at our $R_0$. This sets a strong prior for the analysis of data obtained from global mm-VLBI aiming at resolving Sgr~A* \citep{2000ApJ...528L..13F,2009astro2010S..68D}.

A potential caveat of our analysis might be that the physical model of the orbit is too simple. So far, S2 did not reveal any signs of binarity. For GRAVITY, S2 is an unresolved point source \citep{2017A&A...602A..94G}. The resolution of GRAVITY in GC observations is around $2.2\,\mathrm{mas} \times 4.7\,$mas, excluding a source extension larger than or a companion further than $\approx 1\,$mas. \citet{2018ApJ...854...12C} used S2's radial velocity data and report an upper limit of $M_\mathrm{companion} \sin i \le 1.6 M_\odot$ for periods between 1 and 150 days. Longer periods would not be stable against tidal break-up. Further, the motion of either S2 or Sgr~A* could be affected by yet unknown massive objects in the GC. To some extent such a perturbation can always be absorbed into the orbital elements \citep{2010MNRAS.409.1146G}, resulting in biased estimates for the parameters.
According to our current knowledge, S2 is a suitable probe for $R_0$. It is an ordinary massive main sequence star of type B0 - B3 \citep{2003ApJ...586L.127G, 2008ApJ...672L.119M, 2017ApJ...847..120H}. The atmospheric absorption lines we use are expected to be fair tracers of the motion of the star, together with its (unresolved) photocenter.

The value from \citet{2016ApJ...830...17B}, $R_0 = 7.86 \pm 0.14 \pm 0.04\,$kpc, is discrepant with our result. However, it comes from a combined fit of the stars S2 and S38. The S2-only result of these authors is $R_0 = 8.02 \pm 0.36 \pm 0.04\,$kpc, which is completely consistent with our result. Further we note that combining different stars in the orbit fit tends to change the parameters mass and $R_0$ by rather large amounts \citep{2017ApJ...837...30G}. This is because small inconsistencies in the data sets are amplified by the fact that in the mass-$R_0$ plane two narrow, curved posterior distributions are combined. The statistical error of a combined fit does not catch this and could thus miss part of the true uncertainties. 

Overall, we used accurate radial velocities from SINFONI and proper motions from GRAVITY of the star S2 as it orbits Sgr~A* to set the absolute size of the orbit and determine the distance to the GC with unprecedented accuracy to $R_0 = 8178\,$pc. The statistical error is only $13\,$pc, and is dominated by the measurement errors of the radial velocities. The systematic error of $22\,$pc is dominated by the calibration uncertainties of the astrometry. Our analysis also demonstrates that the relative velocity of the LSR along the line of sight to Sgr~A* is consistent with zero to within few km/s, implying that Sgr~A* is at rest in the GC and the LSR is moving tangentially. 
The addition of further SINFONI and GRAVITY data taken in 2018 also allowed us to increase the significance of the previously published measurement of the gravitational redshift caused by Sgr~A* to $20\sigma$.   

\begin{acknowledgements}
We are very grateful to our funding agencies (MPG, ERC, CNRS, DFG, BMBF, Paris Observatory, Observatoire des Sciences de l'Univers de Grenoble, and the Funda{\c{c}}{\~a}o para a Ci{\^e}ncia e Tecnologia), to ESO and the ESO/Paranal staff, and to the many scientific and technical staff members in our institutions who helped to make NACO, SINFONI, and GRAVITY a reality. S.G. acknowledges support from ERC starting grant No. 306311 (PROGRESO). F.E. and O.P. acknowledge support from ERC synergy grant No. 610058 (BlackHoleCam). J.D., M.B., and A.J.-R. were supported by a Sofja Kovalevskaja award from the Alexander von Humboldt foundation. A.A. and P.G. acknowledge support from FCT-Portugal with reference UID/FIS/00099/2013.
\end{acknowledgements}

%
%

\bibliographystyle{aa} 
\bibliography{r0}

\begin{appendix}
\section{Radial velocities from SINFONI}
\label{appa}

For the SINFONI data we improved the wavelength calibration. Explicitly, we modified our atmospheric line list that serves as reference for the wavelength calibration, by excluding double lines or lines with low SNR following the line atlas of \citet{2000A&A...354.1134R}. We also improved the fine-tuning of the spectrum to the OH lines, leading to an improved wavelength dispersion solution. With this changes we typically achieve a calibration error of below 2 km/s, measured by the residuals of the OH lines used. With the improved data reduction we re-reduced all available data since October 2004. The earlier data (two epochs in 2004 and one in 2003) were obtained during commissioning time and need a dedicated calibration procedure, which we did not repeat. We combined data from different nights when the expected velocity change was smaller than the calibration error. We omitted one measurement from 2008 with low SNR (from a single 10-minute exposure) and included one more epoch from 2009 and 2015 each, and two more from 2010 and 2011 each. We split up data that previously was combined into one cube into two epochs in two occasions, in 2013 and 2015.

For spectra in which both the He-I line ($2.112\,\mu$m) and Brackett-$\gamma$ ($2.166\,\mu$m) lines are unaffected by atmospheric residuals, we used template fitting to determine the radial velocities. For this we fitted the long-time average S2 spectrum \citep{2017ApJ...847..120H} to the data. For spectra with sufficient SNR and no artefacts (as from imperfect atmosphere correction) template fitting yields more accurate velocities. When either of the lines showed artefacts we fitted a double-Voigt profile to the other, unaffected line. 

The errors are a combination of fit error and wavelength calibration uncertainty. The fit error is obtained from the formal fit error $\sigma$, the signal-to-noise ratio (SNR) and by varying the pixel selection. For the SNR-related error we established a relation between $\sigma$ and SNR of $\sigma \propto \mathrm{SNR}^{-0.92}$. The 1 / SNR behaviour is consistent with the uncertainty of a centroid fit \citep{2010MNRAS.401.1177F}. To assess the impact of different background subtractions and extraction regions we extracted eight spectra for each observation and determined the standard deviation of the radial velocities from the different masks. Since these three error estimates are strongly correlated, we used the largest of the three as fit error. We linearly added the wavelength calibration error to obtain a preliminary error. These preliminary errors establish the relative weight of the different radial velocities. Using these we obtained a preliminary orbit fit, which showed that we overestimated the errors, since the residuals around the best preliminary fit are on average 76.8\% of the errors. Thus we rescaled the errors by that factor.

With this improvement of SINFONI analysis we reach an error of $\approx 7\,$km/s for the best data. The median error is $12.3\,$km/s, which is an improvement by 46\% compared to the previous set of radial velocity data.

\section{Astrometry from GRAVITY data}
\label{appb}

\subsection{Data selection}
We started from all observations of Sgr~A* or S2 (793 exposures, each $30 \times 10\mathrm{s} =$ $5\,$min on source, i.e. a total of 66 hours on source), irrespective of observing conditions and instrument performance. 

In 2017, S2 was still at a distance of $54 - 67\,$mas from Sgr~A*, which is comparable to the photometric field of view (FWHM$\approx 65\,$mas), and the exposures pointing on S2 had too little flux from Sgr~A* injected into the fibers for a robust interferometric binary signature \citep{2019arXiv190310937P}. We thus only considered the observations centered on Sgr~A* (261 exposures). We further rejected all Sgr~A* observations for which the instrument internal pupil control \citep{2017A&A...602A..94G} reported an error $>6\,$cm for any of the telescopes (12 exposures), or for which the pointing of any telescope was too far from Sgr~A* (83 exposures). We used a box spanning $\Delta$R.A.$= -45 \,... 10\,$mas, $\Delta$Dec.$= -30\, ... 30\,$mas around Sgr~A*, avoiding especially pointings towards the opposite side of S2. This selection keeps 166 exposures in 2017.
 
For 2018, we had 373 exposures on Sgr~A*. Again, we rejected exposures with pupil errors $>6\,$cm (18 exposures). Because of a newly introduced laser-metrology guiding with substantially improved pointing accuracy, we rejected exposures already when the estimated pointing error for any telescope was outside $\Delta$ R.A. / $\Delta$ Dec.$ = -10\, ... 10\,$mas around Sgr~A* (35 frames). Because S2 was always closer than $23\,$mas to Sgr~A* during our March - June 2018 observing campaigns, both sources were well within the photometric field of view. We also used the 43 exposures centered on S2 obtained during this period. Out of those observations we rejected three exposures because of a pupil error $>6\,$cm, and five exposures because of a pointing error larger $\Delta$ R.A. / $\Delta$ Dec.$ = -10\, ... 10\,$mas. This yields a total of 355 exposures in 2018. 
 
\subsection{Binary fitting and correction for atmospheric refraction}
In a second step, three independent subgroups fitted the individual exposures with a binary model as described in \citet{2018A&A...615L..15G,2018A&A...618L..10G}, using three different codes ("Waisberg (W)", "Pfuhl (P)", "Rodriguez-Coira (R)"). The codes differ in detail in the relative weighting of closure-phases, visibilities, and square visibilities, the free fit parameters (e.g. color of Sgr~A*, flux-ratio per telescope, etc.), and the numerical implementation (e.g. least-square minimization or MCMC), but give overall consistent results for the binary separations. 
 
We further corrected each binary fit for the differential atmospheric refraction between the comparably "blue" S2 and "red" Sgr~A* (see Appendix A7.4 of \citet{2018A&A...618L..10G}). Because Sgr~A* is in its faint, quiescent state for most of our observations, we used the redder, low-flux spectral index $S_\nu \propto \nu^{-1.6}$ from \citet{2018ApJ...863...15W} for the subsequent analysis. With $S_\nu \propto \nu^{2}$ for S2, and for the given effective spectral resolution of $127\,$nm (low resolution mode of GRAVITY), the difference in effective wavelength between S2 and Sgr~A* is $\Delta \lambda = 2.2\,$nm, and the resulting atmospheric differential refraction is $\Delta R = 45\,\mu$as / nm $\times\, \Delta \lambda  \tan z = 99\,\mu$as$ \,\tan z$, where $z$ is the zenith distance. Because we typically observed the GC close to zenith, the atmospheric differential refraction was on average only $30\,\mu$as, and often with opposite signs during a night, therefore resulting in a mean correction of $\Delta$R.A. $= -1\,\mu$as and $\Delta$Dec.$ = -5\,\mu$as.
  
\subsection{Outlier rejection and nightly averaging}
For each of the three sets of binary fits we determined a preliminary orbit for error scaling and outlier rejection. We rejected those observations, for which the residuals were outside the 80\% quantile constructed  in the two-dimensional, error-normalized position residual plane\footnote{The 80\% quantile area is constructed using the Mathematica based Quantile Regression package https://raw.githubusercontent.com/antononcube/MathematicaForPrediction/
master/QuantileRegression.m, Version 1.1, written by Anton Antonov.}. The final data set contains 818 (W), 795 (P), and 737 (R) binary fits, corresponding to about 400 exposures of five minutes each, i.e. about 33 hours on source. We combined these and derived nightly (error-weighted) mean and standard errors (with variance weights). Only in those few cases when we had less than ten binary-fits per night (26/27 March 2017, 28/29 March 2017, 10/11/12 July 2017), we combined several nights to one average. The statistical 1D astrometric error of these combined nightly averages are between $10 - 110\,\mu$as.
   
\subsection{Correction for effective wavelength, systematic error, and final error scaling}
In a last step we corrected the nightly average separation for the effective wavelength shift of $2.3\,$nm (0.1\%) between the wavelength calibration with our $2800\,$K calibration lamp, and the very red, highly dust obscured S2/Sgr~A* data (see Appendix A7.2 in \citet{2018A&A...618L..10G}). 
   
To account for the systematic error in the wavelength calibration, which we estimate to 1/20 detector pixel or equivalent $2.5\,$nm, we added in square the corresponding scale error of 0.11\%. This error in the effective wavelength translated in an astrometric error of about $10\,\mu$as for the time around peri-passage, and up to $\Delta$R.A. $= 66\,\mu$as and $\Delta$Dec. $= 33\,\mu$as for March 2017, when the S2-Sgr~A* distance was largest in our observations. 

Finally, to account for unknown additional errors, we scaled the GRAVITY astrometric errors by a factor 2.2 to match the residuals of a best fitting, preliminary orbit. The resulting astrometric errors around the S2 peri-passage in our data from 24 April - 27 June 2018 are $\Delta$R.A. $= 22 - 101\,\mu$as and $\Delta$Dec. $= 38 - 112\,\mu$as, with a mean of $51\,\mu$as and $60\,\mu$as, respectively.

\section{Systematic error of the GRAVITY astrometry}
\label{appc}
We obtained the GRAVITY astrometry in the single-field mode. S2 and Sgr~A* were close enough in 2017 and 2018 to be fed into the interferometer by a single fiber, the acceptance aperture of which is matched to the telescope point spread function of $\approx 65\,$mas. The two sources appear as an interferometric binary to GRAVITY, which means none of the more complex dual-beam aspects of the instrument \citep{2017A&A...602A..94G} enter in the measurement. 
The standard equation of interferometric astrometry $\Delta \mathrm{OPD} = \vec{s} \times \vec{B}$ sets the effective image scale, where $\vec{B}$ is the baseline and $\vec{s}$ the separation vector one wishes to measure. The accuracy of the interferometric baselines and how well we can measure the OPD thus set the accuracy of $\vec{s}$. 

The value for the baseline length to use is the so-called "imaging baseline" in the sense of \citet{2013ApJ...764..109W,2014A&A...567A..75L}. The telescope position is then defined by the photocenter of the entrance pupil plane appodized by the fiber mode in the pupil plane. While the telescope geometry is known to the mm-level, the active mirrors controlling the fiber mode to pupil overlap are more critical and actually limit the baseline accuracy. A systematic error occurs from how well the fiber mode is aligned with the reference point of the pupil tracker. Also, a vignetting of the pupil would bias the baselines. For an error estimate we use the stability of the pupil position, assuming that the alignment uncertainties overall are at that level. It amounts to  $4\,$cm in the primary mirror space. For the mean baseline length of $81.2\,$m an error of $4\,$cm corresponds to 0.05\% or $4\,$pc on $R_0$.

The wavelength accuracy of the effective wavelengths sets the accuracy of the OPD. From the standard calibrations of GRAVITY we estimate that the wavelength accuracy of the interferogram pixels is 0.11\% or $9\,$pc on $R_0$. It is owed to the faintness of S2 (for interferometric standards), which dictates that we need to observe S2 in low-resolution mode with $R \approx 22$, which corresponds to a wavelength sampling of $50\,$nm/pixel.

When analyzing the results from the three subgroups and fitting codes separately, the standard deviation in the best estimate $R_0$ is $16\,$pc. This takes care of the uncertainty in the binary model fit to the GRAVITY data. The difference between the objective outlier rejection and the manual frame selection of GRAVITY collaboration (2018a) results in a difference in $R_0$ of $15\,$pc. For this estimate, we carried forward the analysis of \citet{2018A&A...615L..15G} with the new data up to the end of 2018, and included the atmospheric refraction effects. This error, however, is not independent of the one from the fitting by subgroups, and we include the larger of the two ($16\,$pc).

The color difference of S2 and Sgr~A* is not known very well, and we include the difference in $R_0$ determined with and without correction of the atmospheric differential dispersion in our error. It amounts to $5\,$pc.

Adding the different contributions in quadrature, we conclude that the total systematic error on the astrometry is $19\,$pc, corresponding to 0.24\%. 

\end{appendix}

\end{document}